\documentclass[useAMS,usenatbib]{mn2e}
\usepackage{times}
\usepackage{graphicx}

\title[A Bayesian analysis of the primordial power spectrum]{A Bayesian analysis of the primordial power spectrum}
\author[M. Bridges et al.]
  {M. Bridges,$^1$\thanks{E-mail: m.bridges@mrao.cam.ac.uk}
   A.N. Lasenby,$^1$ M.P. Hobson$^1$\\
  $^1$Astrophysics Group,
      Cavendish Laboratory, Madingley Road,
      Cambridge CB3 0HE, UK\\
}

\date{Accepted ---. Received ---; in original form \today}
\pagerange{\pageref{firstpage}--\pageref{lastpage}}
\pubyear{2005}

\begin{document}
\label{firstpage}
\maketitle

\begin{abstract}
We present a Bayesian analysis of large-scale structure (LSS) and cosmic microwave background (CMB) data to constrain the form of the primordial
power spectrum. We have extended the usual presumption of a scale invariant spectrum to include: (i) a running spectral index; (ii) a
broken spectrum arising perhaps from an interruption of the potential driving inflation; (iii) a large scale cutoff in power as the first
year WMAP results appear to indicate; (iv) a reconstruction of the spectrum in eight bins in wavenumber; and (v) a spectrum resulting
from a cosmological model proposed by Lasenby \& Doran, which naturally exhibits an exponential drop in power on very large scales. The
result of our complete Bayesian analysis includes not only the posterior probability distribution from which parameter estimates are
inferred but also the Bayesian evidence. This evidence value is greater for a model with fewer parameters unless a more complicated model
provides a significantly better fit to the data, thus allowing a powerful method of model selection. 
We find that those models exhibiting any form of cutoff in power on large scales consistently produce higher evidences than either the
Harrison-Zel'dovich or single spectral index spectra. In particular, within the best-fit concordance cosmology, we find the Lasenby \&
Doran spectrum to show significantly larger evidence as compared to the other models.
\end{abstract}


\section{Introduction}

Recent cosmological surveys of the cosmic microwave background (CMB), in
particular observations by the Wilkinson Microwave Anisotropy Probe (WMAP), and
large-scale structure (LSS) data, such as from the two degree field galaxy redshift survey (2dFGRS), have
provided a unique tool in constraining the structure and contents of the Universe
(\citealt{Spergel}; \citealt{Rebolo}). One of the most exciting results to emerge from these analyses
 is possible structure in the primordial power spectrum, features in which can be used to constrain
early Universe physics, including inflationary models.

The simplest model of inflationary perturbation generation involves a single, slowly rolling inflaton field producing a 
nearly scale-invariant spectrum, although even this is only true in the
limit of an infinitely slow roll with infinite Hubble expansion damping
\citep{Dodelson}. More generally, phase transitions from force unification in the early
Universe \citep{barriga}, multiple field inflation, hybrid inflation \citep{Linde}
and other models (\citealt*{Adams}; \citealt{Wang}) all produce features in the primordial
spectrum. It should also be noted that a scale-invariant spectrum does \emph{not}
uniquely identify inflation as the source of density perturbations, but a
spectrum containing specific features \emph{would} be indicative of a specific
model of inflation. Hence the study of the form of the spectrum is crucial in
analysis of future data.

Generalisations of the shape of the primordial spectrum have been attempted on various physical and
observational grounds to construct an \emph{a priori} parameterisation (e.g.
\citealt{Recon}), while others have attempted to reconstruct an unknown spectrum
directly from the data (\citealt{Wang}; \citealt{Recon}; \citealt*{Silk}; \citealt{Steen}; \citealt{Souradeep}). In this paper
we will mostly employ the latter method, following closely the Bridle et al. analysis despite the obvious disadvantage of weakening
constraints on remaining cosmological parameters and the creation of further
degeneracies by the inclusion of more primordial parameters. One may, however employ a fully Bayesian extension of the Markov-Chain
Monte Carlo (MCMC) method (see Sec. \ref{MCMCsec}) to select the most appropriate model according to its Bayesian evidence 
(see \citealt{Slosar}; \citealt{Beltran}; \citealt{Trotta} and most recently
\citealt*{Parkinson}). This method favours simple models
with fewer parameters over a more elaborate model, unless the latter can provide a
substantially better fit to the data.  

Using a parameterised spectrum requires constraining any analysis to a small sample of the huge
number of possibilities, some of which have been outlined above. Observational features can refine the choice, 
for example the apparently low power observed in the first three CMB multipoles by WMAP. Although some
authors \citep{efstathioua} have suggested that these results are not statistically significant,
in our analysis these data will always prefer a model with low power on these scales. We also
test a broken spectrum caused perhaps by a double field inflationary potential or momentary
pauses in the slow roll of the inflaton \citep{barriga}. \citet{Doran} arrived at
a spectrum naturally incorporating an exponential fall off in power on large scales by considering evolution of closed Universes out of a big bang singularity and with a novel boundary
condition that restricts the total conformal time available in the Universe. The fact that this model features the type
of cutoff that has been suggested on phenomenological grounds from the data \citep{efstathioub} make it an
intriguing possibility.

\section{Model Selection Framework}

\subsection{Markov Chain Monte Carlo sampling}
\label{MCMCsec}
A Bayesian analysis provides a coherent
approach to estimating the values of the parameters, $\mathbf{\Theta}$, and their
errors and a method for determining which model, $M$, best describes the data, $\mathbf{D}$.
Bayes theorem states that
\begin{equation} P(\mathbf{\Theta}|\mathbf{D}, M) =
\frac{P(\mathbf{D}|\mathbf{\Theta},
M)P(\mathbf{\Theta}|M)}{P(\mathbf{D}|M)},
\end{equation}
where $P(\mathbf{\Theta}|\mathbf{D}, M)$ is the posterior,
$P(\mathbf{D}|\mathbf{\Theta}, M)$ the likelihood,
$P(\mathbf{\Theta}|M)$ the prior, $P(\mathbf{D}|M)$ the
Bayesian evidence. Conventionally, the result of a Bayesian parameter
estimation is the posterior probability distribution given by the
product of the likelihood and prior. In addition however, the posterior distribution may be used
to evaluate the Bayesian evidence for the model under consideration.

We will employ a MCMC sampling procedure to
explore the posterior distribution using an adapted version
of the \emph{cosmoMC} package \citep{cosmomc} with four CMB
datasets; WMAP (\citealt{WMAP}; \citealt{WMAP2}; \citealt{WMAP3}), ACBAR
\citep{ACBAR} the latest VSA results \citep{VSA} and CBI \citep{CBI}. We also include
the 2dF Galaxy Redshift Survey \citep{2dF}, the Sloan Digital Sky Survey \citep{Abazajian} and the
\emph{Hubble Space Telescope (HST)} key project \cite{Freedman}. 
In addition to the
primordial spectrum parameters, we parameterise each model using
the following five cosmological parameters; the physical baryon density $\Omega_b
h^2$; the physical cold dark matter density $\Omega_{c} h^2$; the total energy density
$\Omega_0$ (parameterised as $\Omega_k = 1 - \Omega_0$); the Hubble parameter $h$ ($H_0 = h \times100 \mbox{kms}^{-1}$) and the redshift of
re-ionisation $z_{re}$.

\subsection{Bayesian evidence and simulated annealing}
\label{bayes}
The Bayesian evidence can be
defined as the average likelihood over the entire parameter space of the model: 
\begin{equation}
\int{P(\mathbf{D}|\mathbf{\Theta}, M)P(\mathbf{\Theta})}d^N\mathbf{\Theta}.
\end{equation}
In general models with fewer parameters and hence small parameter spaces will have larger
evidences.
Naively one could evaluate the evidence by sampling from the posterior
distribution function randomly covering the entire parameter space
and simply find the average likelihood value. In practise however the value of the likelihood at its maximum is typically many orders
of magnitude larger than that at any extremes of the space and so
very many samples would need to be taken away from the peak in order
for their inclusion to make any appreciable difference. To overcome this one can use the
numerical method of simulated annealing, also known as thermodynamic integration, to slowly increase the height of the
peak in the likelihood
relative to the surrounding background value. Using a random sampling of the space, such
as MCMC, one can ensure that all of the parameter space, not just the area around the
peak likelihood is explored.  

Using such an annealing schedule we draw samples from
$P(\mathbf{D}|\mathbf{\Theta},M)^{\lambda}P(\mathbf{\Theta}|M)$
where $\lambda$ is the inverse temperature and is raised from
$\approx 0$ to $1$. One can then define the evidence as a function of $\lambda$ as
\begin{equation}
E(\lambda) = \int P(\mathbf{D}|\mathbf{\Theta},M)^{\lambda}P(\mathbf{\Theta}|M) d^N \mathbf{\Theta}.
\end{equation}
We require that the priors be normalised to unity over the parameter space considered
thus  $E(\lambda=0)=1$ and we will also find it computationally simpler to work in the
natural logarithm of evidences, therefore:
\begin{eqnarray*}
\ln E(\lambda) &=& \ln E(0) + \int_0^1 \frac{d \ln E(\lambda)}{d \lambda}d \lambda \\
&=& \int_0^1 \frac{d \ln E(\lambda)}{d \lambda}d \lambda \\
&=& \int_0^1 \frac{1}{E}\frac{dE(\lambda)}{d \lambda} d \lambda.
\end{eqnarray*}
Performing the derivative of $E(\lambda)$ gives
\begin{equation}
\ln E(\lambda) = \int_0^1\frac{\int P(\mathbf{D}|\mathbf{\Theta},M)^{\lambda}
\ln P(\mathbf{D}|\mathbf{\Theta},M)P(\mathbf{\Theta}|M) d^N \mathbf{\Theta}}{\int P(\mathbf{D}|\mathbf{\Theta},M)^{\lambda}
P(\mathbf{\Theta}|M) d^N \mathbf{\Theta}}d\lambda.
\end{equation}
This gives the expectation value of $\ln P(\mathbf{D}|\mathbf{\Theta},M)$
over the parameter space defined by the extent of the priors. Thus we can simply
sum over these values at each step $\Delta \lambda$ in the annealing schedule and
divide by the total number of samples to find the log evidence value
\begin{equation}
\ln E(\lambda) \approx \frac{1}{N_\lambda} \sum_{i=1}^{N_\lambda}
\ln P(\mathbf{D}|\mathbf{\Theta},M)_i.
\label{evidence}
\end{equation}
The evaluation of this integral was performed using a reverse annealing schedule (as set
out in \citet{Beltran}) where
$\lambda$ is lowered from $1$ to $0$ once the Markov chain has found its stationary
point through a period of normal burn-in, which typically takes $< 500$ samples. The
number of annealing steps $N$ is not a chosen constant but is instead determined by
the stopping criterion that the final log evidence value would not change by more than a
given threshold (empirically set to 0.001). Accordingly $\lambda$ at each step is given
as:  
\begin{equation}
\lambda(N) = (1-\epsilon)^N.
\end{equation}
where $\epsilon$ is a user defined parameter of order $5\times10^{-5}$. With this method
log evidence values varying to within only one unit were successfully obtained from independent
chains. The total number of sampler calls made during the evidence burn in was
$\approx 3000$ deliberately large to ensure the chain was at a stationary point. During the reverse annealing schedule typically $\approx 15000$ calls were made.

\section{Primordial Power Spectrum Parameterisation}
\label{parameterisation}
The correlation function $\xi$ over a range $r$ of a density field,
such as the matter density field in the Universe is defined as the
product of the density contrast at two separate points, $\mathbf{x}$ and
$\mathbf{x+r}$
\begin{equation}
\xi(r)\equiv\langle\delta(\mathbf{x})\delta(\mathbf{x+r})\rangle.
\end{equation}
The power spectrum $P(k)$ is simply the inverse Fourier transform of
the correlation function or the ensemble average power: 
\begin{equation}
P(k)\equiv\langle|\delta_k|^2\rangle.
\end{equation}
The basic inflationary paradigm inflates quantum zero-point oscillations of the inflaton to macroscopic scales,
providing a source of density fluctuations which would then evolve via gravitational
collapse to form the observed large-scale structure today. The power spectrum of these
primordial fluctuations can then provide unique constraints on the dynamics of the inflationary epoch.
The homogeneity of the Universe on large scales suggests the simplest spectrum and most 
obvious first candidate is scale invariant. This intuition is confirmed by the results of slow-roll inflation,
predicting the slightly sloped power law spectrum 
\begin{equation}
P(k)=A k^{n-1}.
\label{single_index}
\end{equation}
where $\mid n-1 \mid \ll 1$. 
It can be shown that the Fourier components $\delta_k$ are uncorrelated and have random phases
meaning the power spectrum encodes all of the stochastic properties of the density field.

\subsection{Harrison-Zel'dovich and Power-Law Parameterisations}
\label{power-law} 
The simplest scale invariant power spectrum can be parameterised with one free parameter, an amplitude; $A_s$ as:
\begin{equation}
P(k) = A_s,
\end{equation}
This is known as the Harrison-Zel'dovich (H-Z) spectrum. Slow roll inflationary scenarios would be expected to imprint a slight slope, leading
to the familiar power law form 
characterised by a spectral index $n_s$ (see Eqn. \ref{single_index})
A link between $n_s$ and the potential driving inflation is well
established \citep{Lyth} but the spectrum is only an exact power law for an
exponential inflationary potential, so in general the spectral index should be some function
of scale $n(k)$. Therefore we can also characterise the form of the spectrum as
\begin{equation}
P(k) = A_s
\left(\frac{k}{k_0}\right)^{n-1+(1/2)\ln (k/k_0)(dn/d\ln k+\cdot\cdot\cdot)}.
\end{equation}
where $dn/d\ln k$ is the running parameter $n_{run}$.
For most standard models we would expect $n_{run} \approx 0$. The entire
spectrum pivots about a scale $k_0$ which is also the position in $k$-space at which the amplitude is set. In
keeping with 
previous studies we have set this scale to $0.05\mbox{Mpc}^{-1}$. 

\begin{figure}
\begin{center}
\includegraphics[width=\linewidth]{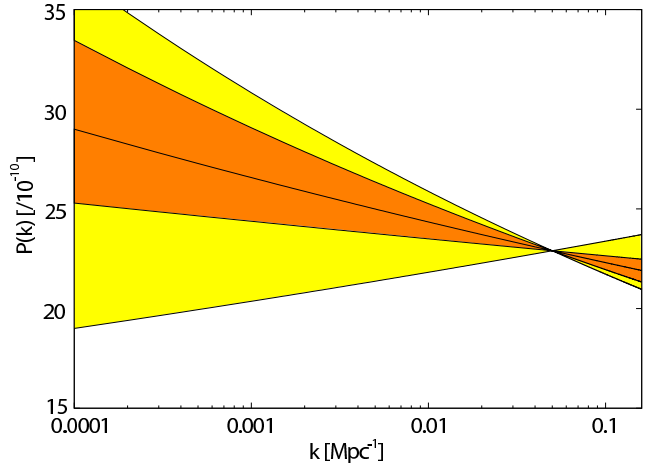}
\caption{Best fit single spectral index power law spectrum with $1\sigma$ and $2\sigma$ errors (shown with shaded areas).[Note that, for
clarity, the error in $A_s$ is not shown]}
\label{lambda}
\end{center}
\end{figure}

\begin{figure}
\begin{center}
\includegraphics[width=\linewidth]{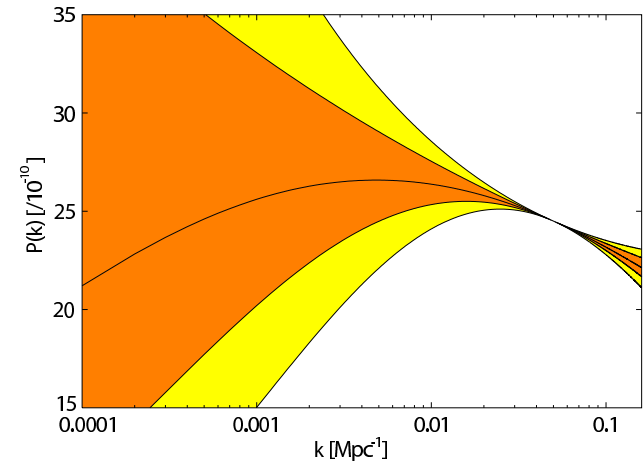}
\caption{Best fit running spectral index power law spectrum with $1\sigma$ and $2\sigma$ errors (shown with shaded areas).[Note that, for
clarity, the errors in $A_s$ and $n_s$ are not shown]}
\label{run}
\end{center}
\end{figure}

The complete parameter estimates of our analysis using the data described above are shown in Table \ref{MCMC}. The inclusion of the possibility of a non-flat geometry
lowers the value of $h$ considerably, a point to which we will return in Sec. \ref{doran}. We find a best-fit single index spectrum (see
Fig. \ref{lambda})
that, despite the inclusion of more recent data and extension to non-flat geometries, does not differ appreciably from the best-fit index
found by \citet{Spergel}. The inclusion of a running index (see Fig. \ref{run}) weakens the constraint slightly to $0.93\pm0.05$. 
To $1\sigma$ level the constraint on $n_{run}$ provides only slight evidence for a dependence of $n_s$ on $k$. 

\begin{table*}
\begin{center}
\caption{MCMC parameter constraints for H-Z, single index and running
index parameterisations (mean $\pm 1 \sigma$ errors)}
\begin{tabular}{|c||c||c||c||c|}
    \hline
 Parameter &  H-Z & Single Index & Running Index & Priors\\ 
    \hline
 $n_s$ & - & 0.96 $\pm$ 0.02 & 0.93 $\pm$ 0.05& [0.5,1.5]\\
 $n_{run}$ & - & - & -0.034 $\pm$ 0.032& [-0.15,0.15]\\
 $A_s(\times 10^{-8})$ &  27.1 $\pm$ 1.1 & 22.6 $\pm$ 1.1  & 24.9 $\pm$ 1.2 &[14.9,54.6]\\
 $\Omega_b h^2$ &  0.0240 $\pm$ 0.0006   & 0.0229 $\pm$ 0.0009 & 0.0225 $\pm$ 0.0012& [0.005,0.1]\\
 $\Omega_c h^2$ &  0.116 $\pm$ 0.009   & 0.118 $\pm$ 0.008  &  0.125 $\pm$ 0.012& [0.01,0.99]\\
 $\Omega_k$ &    $-0.026^{+0.018}_{-0.019}$   &$-0.024_{-0.019}^{+0.018}$  &  $-0.022_{-0.019}^{+0.020}$& [-0.25,0.25]\\
 $h$ &      $0.63_{-0.06}^{+0.05}$    &0.61 $\pm$ 0.05   & 0.61 $\pm$ 0.05 & [0.4,1]\\
 $z_{re}$ &   19.0  $\pm$ 2.6    &  13.4 $\pm$ 4.2  & 16.4 $\pm$ 5.5 & [4,30]\\
    \hline
\end{tabular}
\label{MCMC}
\end{center}
\end{table*}

\subsection{Large Scale Cutoff} WMAP observations have pointed to lower than expected power on large
scales. Although the statistical significance of these data points have
been questioned \citep{efstathioub}, currently models with a cutoff in power should be preferred. We did
not attempt to model the form of the spectrum near the
cutoff, instead parameterising the scale at which the power drops to zero,
$k_c$, with a prior of $[0.0,0.0006]$ Mpc$^{-1}$: 
\begin{equation}
P(k) = \left\{ \begin{array}{ll}
         0,& \mbox{$k < k_c$}\\
     A_s \left(\frac{k}{k_0} \right)^{n-1},& \mbox{$k \geq k_c$}\end{array}\right.
\end{equation}
It is worth pointing out however that inherent
cosmic variance limitations at this scale would make constraints of any form difficult to obtain.

\begin{figure}
\begin{center}
\scalebox{0.3}{\includegraphics{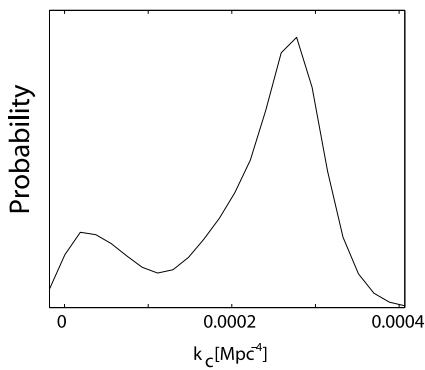}}
\caption{Marginalised histogram of the cutoff in power on large scales}
\label{cutoff}
\end{center}
\end{figure}

On small scales this spectrum behaves just as the single index power law and so constraints on the cosmological
parameters remain essentially unchanged.
A cutoff is preferred
(see Fig. \ref{cutoff}) but the single index model is still not ruled out, illustrated by the non zero value of the likelihood function at $k_c = 0$.
The likelihood function peaks at a cutoff scale of $2.7\times10^{-4}$ Mpc$^{-1}$ and
drops off markedly thereafter towards higher $k_c$. This figure reproduces well
the results of \citet{Recon} and despite a different
mathematical form for the cutoff gives a similar likelihood distribution to
\citet*{Niarchou}, it is also comfortably within the $2\sigma$ upper limit of $k_c < 7.4 \times 10^{-4}$ Mpc$^{-1}$ found by \citet*{Contaldi}
 using an exponential cutoff . 

\subsection{Broken Power Spectrum} Rapid cooling of the Universe during inflation
can result in spontaneous symmetry breaking phase transitions which could
interrupt the potential driving inflation leading to one or more sudden departures
in scale invariance lasting $\approx1 e$ fold \citep{barriga}. When this occurs, the mass of the inflaton field changes suddenly,
 thus interrupting inflation. Although predicting the form of this
interruption is not trivial, it is clear we should observe a drop in power, i.e.
a break in the primordial power spectrum, after which the spectrum should return to
scale invariance. We will assume the spectrum has the general form
\begin{equation}
P(k) = \left\{ \begin{array}{ll}
         A,& \mbox{$k \leq k_s$}\\
     Ck^{\alpha-1},& \mbox{$k_s < k \leq k_e$}\\
        B,& \mbox{$k > k_e$}\end{array} \right.
\end{equation}
where the values of $C$ and $\alpha$ are chosen to ensure continuity. Four power
spectrum parameters were varied in this model: the ratio of amplitudes before and after the
break $A/B$ with prior $[0.3,7.2]$; $k_s$ indicating the start of the break with prior
$[0.01,0.1]$ Mpc$^{-1}$; $\ln(k_e/k_s)$ to constrain the length of the break with prior $[0,4]$ and
normalisation $A$ with prior $[14.9,54.6]\times 10^{-8}$. We also placed a prior that $k_e$ could not exceed 0.1 Mpc$^{-1}$.

\begin{center}
\begin{figure}
\includegraphics[width=\linewidth]{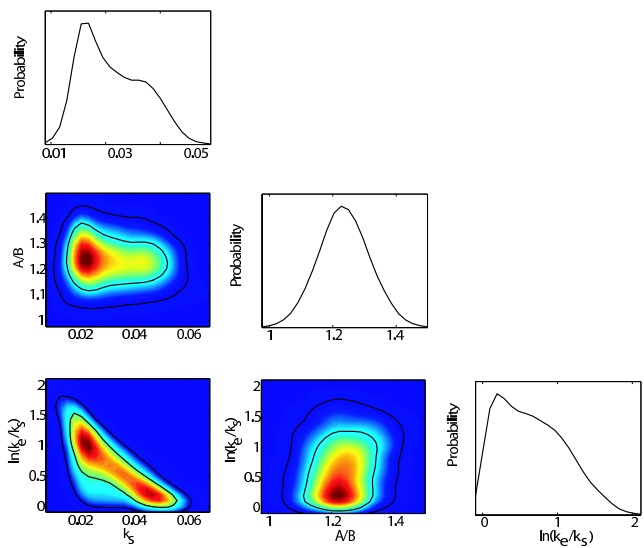}
\caption{Marginalised 1D and 2D probability constraints on $k_s$ $\ln(k_e/k_s)$ and $A/B$,
2D constraints plotted with $1\sigma$ and $2\sigma$ confidence contours.}
\label{barigga}
\end{figure}
\end{center}

\begin{center}
\begin{figure}
\includegraphics[width=\linewidth]{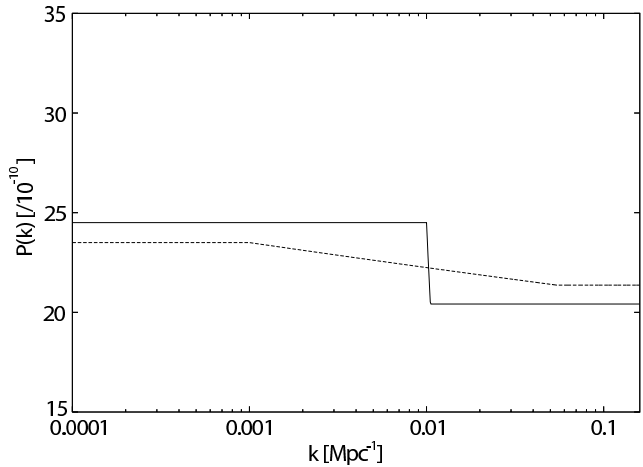}
\caption{Examples of two broken spectra; with an extended and sudden transition in power.}
\label{broken_spectra}
\end{figure}
\end{center}

Constraints on these parameters are shown in Fig. \ref{barigga}, in very good agreement with the Bridle et al.
analysis showing that this
parameterisation is robust in the extension to closed models. From the preference for
a spectral index lower than unity in Sec \ref{power-law} it is clear that a drop in power would
be preferred by the data, especially if the transition was smooth and extended as this
would mimic a tilted spectrum. Indeed this effect is seen by the large peak in the
likelihood surface at very low values of $\mbox{ln}(k_e/k_s)$ (i.e. a large difference
in $k_e$ and $k_s$ implying an extended break). More interesting however is the preference for
a sudden drop on large scales $k\approx0.025$Mpc$^{-1}$, this could be evidence for a phase
transition in the early Universe, or as Bridle et al. point out, could be an artifact of
the combination of WMAP and 2dF datasets. Example spectra featuring both extended and sharp breaks are shown in Fig. \ref{broken_spectra}. 

\subsection{Power Spectrum Reconstruction} 
\begin{center}
\begin{figure}
\includegraphics[width=\linewidth]{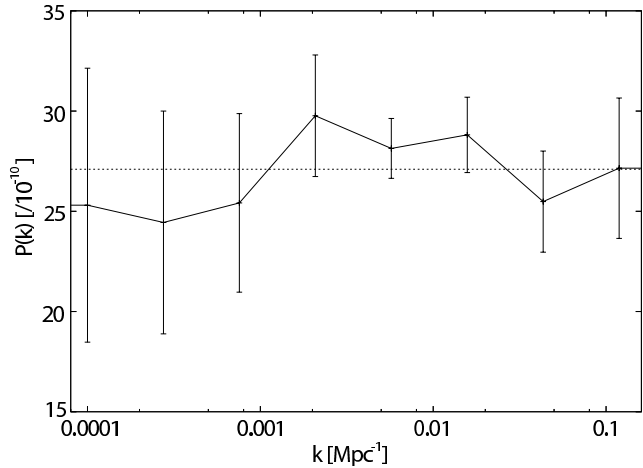}
  \caption{Reconstruction of the primordial power spectrum in 8 bands of $k$, compared to the best fit H-Z spectrum (dotted line).}
\label{binned_spectra}
\end{figure}
\end{center}

The data that can now be used to
constrain the primordial power spectrum are generally analysed in the framework of a specific
model, as we have done up to now. It is possible however that certain features of the
observational data are being overlooked with this method. In order to reveal any such
features we have divided the spectrum into eight bins in wavenumber $k$ and
allowed the amplitude in each bin to vary over the same range as the single amplitude $A_s$ for
the power law spectrum. Instead of using simple `tophat' bin amplitudes we used the same approach as \citet{Recon}, 
linearly interpolating between the bins in $k$. Since current interest has concentrated on large scale departures from
scale invariance we confined the study to bins, of amplitude $a_n$, between 0.0001 and 0.11 Mpc$^{-1}$
parameterised logarithmically with $k_{i+1} = 2.75k_i$ so that
\begin{equation} P(k) = \left\{
\begin{array}{ll} \frac{(k_{i+1} - k)a_i + (k-k_i)a_{i+1}}{k_{i+1}-k_i},& \mbox{$k_i <k <
k_{i+1}$}\\ a_n,& \mbox{$k \geq k_n$}\end{array}\right. 
\end{equation}
We show in Fig. \ref{binned_spectra} the reconstructed
spectrum relative to the best fit Harrison-Zel'dovich (H-Z). The effect of degeneracies between neighbouring bins is minimised with the use of \emph{cosmoMC}'s new parameterisation to explore the $A_s - \tau$ degeneracy, however there is still
considerable variability at the $1\sigma$ level. Our results are mostly in agreement with a H-Z spectrum, although we also find the feature
observed at around 0.06 Mpc$^{-1}$ by \citet{Recon}.
We observe a slight decrement in the power on scales below ~ 0.001 Mpc$^{-1}$, confirming results by \citet{Steen} and \citet{Souradeep}, this effect would
produce a similar decrement in the CMB spectrum at low $l$.  

\subsection{Closed Universe Inflation} 
\label{doran}
\citet{Doran} arrived at a novel model spectrum by considering a boundary condition that
restricts the total conformal time available in the Universe, and requires a closed
geometry. The resultant predicted perturbation spectrum encouragingly contains an exponential cutoff (as
previously suggested phenomenologically by \citealt{efstathioua}) at low $k$ which yields a corresponding deficit in power in the CMB
power spectrum. The shape of the derived spectrum was parameterised by the function:
\begin{equation}
P(k) = A (1-0.023y)^2(1-\exp(-(y+0.93)/0.47))^2,
\end{equation}
where $y=\ln\left(\frac{k}{H_0/100}\times 3 \times 10^3\right)> -0.93$.
\begin{figure}
\begin{center}
\includegraphics[width=\linewidth]{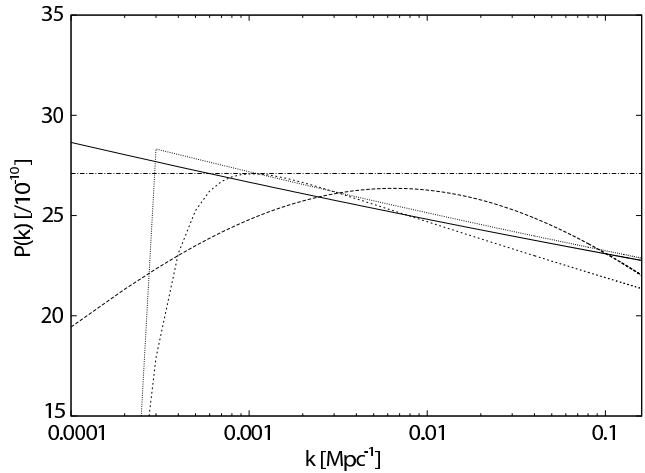}
\caption{Lasenby \& Doran spectrum (short-dashed) shown with best fitting H-Z (dot-dashed), single index with a cutoff
(dotted), without (full) and a running index (long-dashed).}
\label{anthony}
\end{center}
\end{figure}

Unfortunately it has not been possible to produce a routine that can recalculate the form of this spectrum (see
Fig. \ref{anthony}) in non-primordial parameter space fast enough to perform a full MCMC analysis.
The authors did, however suggest a best-fit cosmology
($\Omega_{b}h^2=0.0224$, $\Omega_{cdm}h^2=0.110$, $h=0.6$ and a total energy density $\Omega_{0}= 1.04$) from which we could fit the
primordial spectrum using just an amplitude $A$ with prior $[14.9,54.6]\times 10^{-8}$. Although such a low Hubble parameter would have been ruled out from
previous analyses \citep{Spergel} the extension to non-flat geometries weakens the constraint
considerably, as discussed earlier. The best fitting 
spectrum ($A$ = 29.82 $\pm$ 0.19) is shown with the H-Z, single index, cutoff and running index models for comparison in Fig \ref{anthony}. 
Encouragingly the `turn-over' scale of the L+D spectrum is close to that found in the reconstructed spectrum at about $0.001$ Mpc$^{-1}$,
while at large $k$ the spectrum successfully mimics a power law spectrum with spectral index $n_s\approx0.96$.

\section{Application to Simulated Data}
In order to test the model selection algorithm we produced a series of simulated data sets with known cosmological
parameters and a particular primordial power spectrum parameterisation. We produced two sets: the first intended to represent 
the quality expected from the
forthcoming Planck satellite mission, the second a combination of simulated WMAP first year CMB data and Sloan
Digital Sky Survey (SDSS) LSS data. 
At least eight separate Markov chains were produced giving eight estimates of the evidence and an associated variance. The chosen model was flat with $\Omega_{b}h^2=0.022$,
$\Omega_{cdm}h^2=0.120$, $h=0.72$, optical depth to re ionisation of $\tau = 0.15$ and with primordial parameters $n_s = 0.97$
and
 $A_s = 2.5\times10^{-9}$.

\subsection{Simulated Planck Data} 
\begin{figure}
\begin{center}
\includegraphics[width=\linewidth]{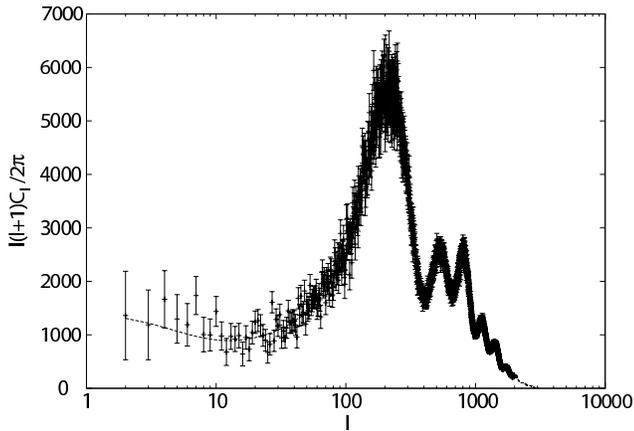}
\caption{Simulated cosmic variance limited data to $l=2000$ (convolved with Gaussian noise) and the model from which it was produced (dashed line).}
\label{planck}
\end{center}
\end{figure}

\begin{figure}
\begin{center}
\includegraphics[width=\linewidth]{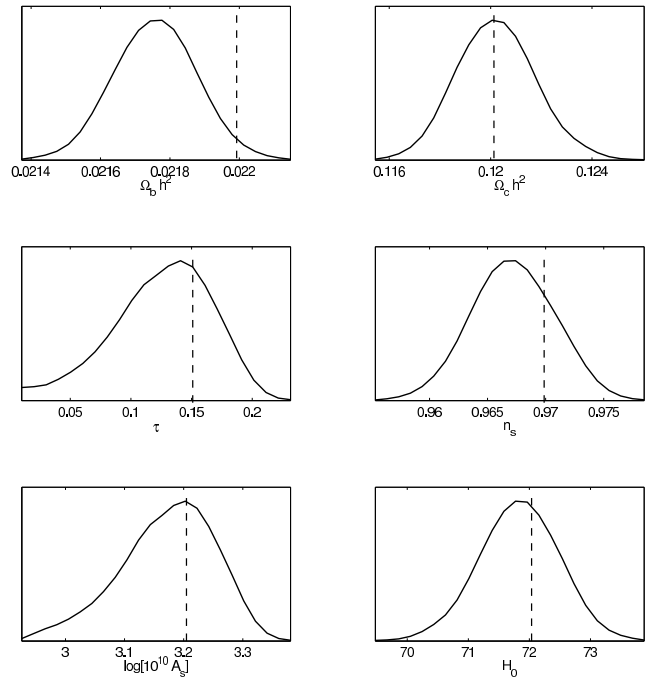}
\caption{Recovered MCMC marginalised parameter constraints using cosmic variance limited data.}
\label{planck_params}
\end{center}
\end{figure}

We approximated mock Planck data as being cosmic variance limited up to $l$ of $2000$, which was sufficiently accurate 
to determine the spectral index $n_s$ in such a
parameterisation to an accuracy of better than $2\%$ and was produced using $C_l$s from the power spectrum
generator \emph{CAMB} (\citep*{camb}) to which we added Gaussian noise with a cosmic variance standard deviation of
\begin{equation}
\Delta C_l^2 = \frac{2}{l(l+1)}C_l^2.
\end{equation}
The CMB spectrum of the chosen model is shown in Fig. \ref{planck} along with the data points used including simulated noise.
Parameter constraints confirm that the sampler is capable of extracting this model from the dataset (see Fig. \ref{planck_params}).

\begin{table}
\begin{center}
\caption{Differences of log evidences for four parameterisations using cosmic variance limited data.}
\begin{tabular}{|c||c|}
    \hline
 \textbf{Model} &  $\mbox{ln}E_{\Lambda}-\mbox{ln}E$ \\
    \hline
 Constant $n$ & 0.0 $\pm$ 2.4\\
 Running &  -4.7 $\pm$ 2.2\\
 Cutoff &  -1.1 $\pm$ 2.9\\
 Broken &  -36.7 $\pm$ 3.3\\
    \hline
\end{tabular}
\label{table2}
\end{center}
\end{table}
Since the absolute evidence value can only be compared between models using the same dataset it is conventional to quote ratios with
respect to a reference model, which in this case will be the single spectral index. Our results (shown in Table \ref{table2})
successfully select the correct model, although with high variance between chains. In particular, the cutoff model is, within
estimated $1\sigma$ error, a very close second. This is not too surprising given that the cutoff spectrum is simply a single index
spectrum above the appropriate cutoff wave vector $k_c$, at which scale cosmic variance is large enough to prevent one obtaining strong constraints. 
The question of what difference in log evidence one requires to say with some certainty which model the data prefers is not a trivial one
and depends to some extend on `intuition'. A
useful guide has been given by \citet{Jeffreys} where a log evidence difference:
$\Delta\mbox{ln} E < 1$ is not significant, $1 < \Delta\mbox{ln} E < 2.5$ significant, $2.5 < \Delta\mbox{ln} E < 5$ strong and $\Delta\mbox{ln} E > 5$ decisive. 
Using this criterion 
we can correctly rule out, with some confidence, the running index spectrum and completely discount the broken model. The combination of
some LSS data sets would doubtless allow a distinction between the cutoff and single index spectrum though
this test has not, as yet been performed.  

\subsection{Simulated WMAP \& SDSS Data}
\begin{figure}
\begin{center}
\includegraphics[width=\linewidth]{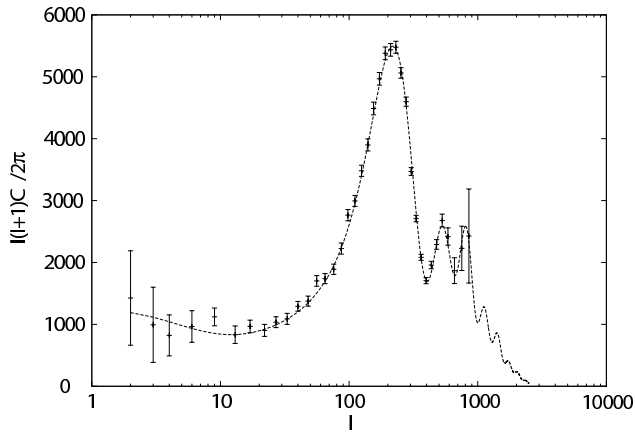}
\caption{Simulated first year WMAP data to $l=856$ and the model from which it was produced (dashed line).}
\label{wmap_mock}
\end{center}
\end{figure}

\begin{figure}
\begin{center}
\includegraphics[width=\linewidth]{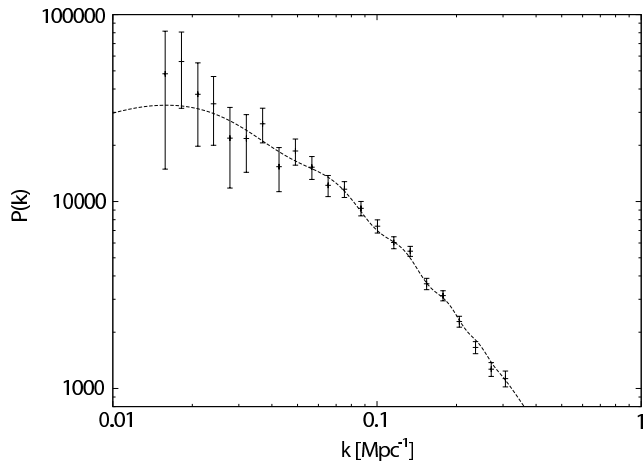}
\caption{Simulated SDSS data and the matter power spectrum from which it was produced (dashed line)}
\label{sdss_mock}
\end{center}
\end{figure}

\begin{figure}
\begin{center}
\includegraphics[width=\linewidth]{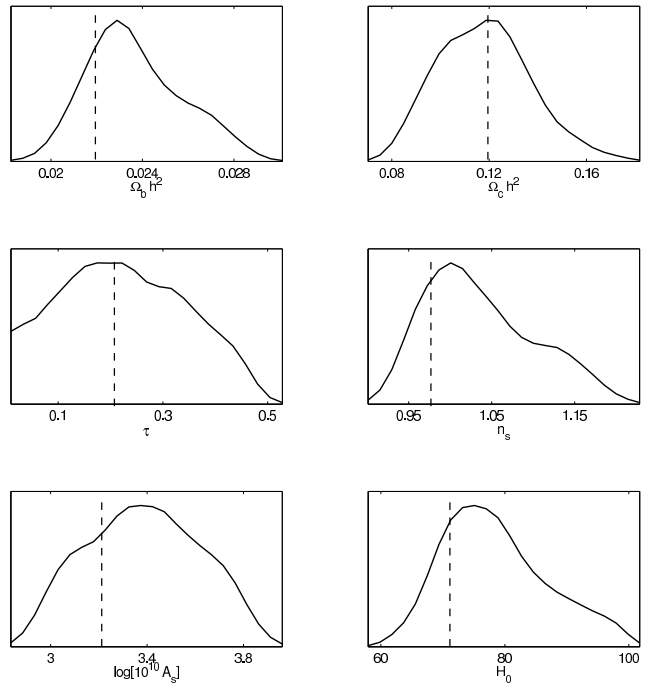}
\caption{Recovered MCMC marginalised parameter constraints for the chosen model using simulated first year WMAP + SDSS data.}
\label{wmap_sdss_params}
\end{center}
\end{figure}

Two further datasets were created to mimic the current data used in our full analysis: a simulated first year WMAP
CMB power spectrum and large-scale structure (LSS) data similar to the SDSS matter power spectrum. Both
spectra were again produced using the \emph{CAMB} generator using the same single index model chosen above. The mock WMAP $C_l$s were
created by adding Gaussian noise using the calculated WMAP errors at multipoles between $2$ and $856$ corresponding to the binning used by the
WMAP team (see Fig. \ref{wmap_mock}).
SDSS data was produced (see Fig. \ref{sdss_mock}) using the binning in $k$ space and calculated errors used by \citep{sloan}.
As is well known the combination of LSS and CMB data is capable of breaking a number of degeneracies in
parameter space and can produce tighter constraints than with CMB data alone (see Fig. \ref{wmap_sdss_params}).
Although differences in the evidence estimates made with this realistic data are somewhat smaller than with simulated Planck data it is still possible to determine the correct model, though not
decisively (Table \ref{wmap_ev}) according to Jeffreys criteria.    
\begin{table}
\begin{center}
\caption{Differences of log evidences of four models using a combination of simulated WMAP and SDSS data.}
\begin{tabular}{|c||c|}
    \hline
 \textbf{Model} &  $\mbox{ln}E_{\Lambda}-\mbox{ln}E$ \\
    \hline
 Constant $n$ & 0.0 $\pm$ 0.8\\
 Running &  -2.3 $\pm$ 0.8\\
 Cutoff &  -1.2 $\pm$ 0.6\\
 Broken &  -1.1 $\pm$ 1.5\\
    \hline
\end{tabular}
\label{wmap_ev}
\end{center}
\end{table}

\section{Application to Real Data}
Given the success of model recovery using simulated data, we can now turn back to real data with some
confidence. All of the parameter estimations in Sec. \ref{parameterisation} were accompanied by at least eight
separate log evidence estimates. We can divide
this part of the analysis into two main sections: firstly a full parameter space exploration for the H-Z, single index,
running, cutoff, broken and binned models; and secondly estimations made only with the reduced primordial parameter
space (i.e. only those parameters affecting the form of the initial power spectrum) within a predetermined
cosmology, to include the Lasenby \& Doran model. This division was necessary because of the restrictions in the form of the  Lasenby
\& Doran spectrum discussed in Sec. \ref{doran}.

\subsection{Full cosmological parameter space exploration} 
  
Statistical uncertainty dominates the results of the full parameter space study (see Table \ref{table4}), leaving it difficult to draw any
meaningful conclusions. All of the central mean values, with the exception of the binned spectrum, are found to differ by less than a
unit in log evidence, which, according to Jeffreys criteria, provides an insignificant difference in evidences. However ranking these
mean values we find that the cutoff spectrum is marginally preferred overall. Although a better chi-squared fit was found by
\citet{Spergel} for a running parameterisation we find no conclusive evidence in its favour in this study. Indeed it is worth noting, any
possible preference for a typical running spectrum($n_{run} \approx -0.035$) could possibly be due to its large scale power decrement
rather than any running. The reconstructed spectrum, revealing any possible structure would be expected to provide the best possible
model fit --however the large increase in parameter space has clearly had a detrimental effect producing a `significant' evidence result
disfavouring the model. In order to reduce the statistical uncertainty in these estimates to the point of distinguishing between the best
fit models we would require of order 50 seperate runs for each parameterisation which at present is not computationally feasible. However
the use of the alternative method of nested sampling (see \citealt{Parkinson}) could allow the reduction of uncertainties to the required
level in a future study.

\subsection{Primordial parameter space exploration}
The Lasenby \& Doran spectrum had previously been calculated for a cosmological model described by
$\Omega_0 = 1.04, \Omega_b h^2 =
0.0224, h = 0.6, \Omega_{cdm} h^2 = 0.110$, and provided a good fit to WMAP and higher resolution data. We
 therefore used this cosmology as a starting point.
Again the results (see Table \ref{table5}) have been normalised to the single index spectrum, and the similarity in evidences make it difficult to distinguish between the
H-Z, single index and cutoff. We can however clearly rule out the running spectrum in this case with some certainty.
Though not a convincing result in this case, the mean evidence of the Lasenby \& Doran spectrum lies slightly higher, quantitatively
confirming the `hand' fit used in \citet{Doran}.

With this relative success for a restricted cosmology, it made sense to
consider the evidence rankings within the best fit concordance cosmology, in
this case the parameter estimates found from the full MCMC simulations of the
power spectrum. A general package for predicting the power spectrum for
arbitrary parameters within the Lasenby \& Doran model is not currently
available, however we know the form to be fairly stable with varying
cosmological parameters. Hence we used the $\Omega_0=1.04$ model described
above as a template to compare with the results from the full MCMC fitting,
which gave parameters $\Omega_0 = 1.024, \Omega_b h^2 = 0.0229, h = 0.61, 
\Omega_{cdm} h^2 = 0.118$. Although these correspond to a different cosmology
from that used for the Lasenby \&  Doran case, the cosmological parameters do
not deviate greatly from these across the different parameterisions, in
particular the cutoff and running cases, hence we feel an evidence comparison
is justified. This result (Table \ref{table6}), shows the L+D model to lie roughly 3 units
of log evidence ahead of the cutoff case, yielding an evidence ratio of roughly
20, which according to the Jeffreys ranking system provides a `strong' model
selection. 

\begin{table}
\begin{center}
\caption{Differences of log evidences with respect to single index model using the full cosmological parameter space.}
\begin{tabular}{|c||c|}
    \hline
 \textbf{Model} &  $\mbox{ln}E_{\Lambda}-\mbox{ln}E$ \\
    \hline
 Constant $n$ & 0.0 $\pm$ 0.7\\
 H-Z    &    0.4  $\pm$  0.8 \\
 Running &  0.5 $\pm$ 0.7\\
 Cutoff &  0.8 $\pm$ 0.8\\
 Broken &  -0.3 $\pm$ 0.8\\
 Binned &   -1.8  $\pm$ 0.7\\
    \hline
\end{tabular}
\label{table4}
\end{center}
\end{table}

\begin{table}
\begin{center}
\caption{Differences of log evidences (for primordial parameters) for all models with respect to single index model within
cosmology: $\Omega_0 = 1.04, \Omega_b h^2 = 0.0224, H_0 = 60, \Omega_{cdm} h^2 = 0.110$.}
\begin{tabular}{|c||c|}
    \hline
 \textbf{Model} &  $\mbox{ln}E_{\Lambda}-\mbox{ln}E$ \\
    \hline
 Constant $n$ & 0.0 $\pm$ 0.6\\
 H-Z    &   -0.4   $\pm$ 0.5  \\
 Running &  -2.1 $\pm$ 0.5\\
 Cutoff &  0.2 $\pm$ 0.6\\
 Broken &  -0.5 $\pm$ 0.7\\
 Binned &  -5.2 $\pm$ 0.7 \\
  Lasenby \& Doran & 0.9 $\pm$ 0.6\\
    \hline
\end{tabular}
\label{table5}
\end{center}
\end{table}

\begin{table}
\begin{center}
\caption{Differences of log evidences (for primordial parameters) for all models with respect to single index model 
within the current concordance cosmology: $\Omega_0 = 1.024, \Omega_b h^2 = 0.0229, h = 0.61, \Omega_{cdm} h^2 = 0.118$, 
as compared to the Lasenby \& Doran model (with parameters as in the previous table)}
\begin{tabular}{|c||c|}
    \hline
 \textbf{Model} &  $\mbox{ln}E_{\Lambda}-\mbox{ln}E$ \\
    \hline
 Constant $n$ & 0.0 $\pm$ 0.5\\
 H-Z    &   -4.4   $\pm$ 0.5  \\
 Running &  -0.8 $\pm$ 0.6\\
 Cutoff &  0.4 $\pm$ 0.5\\
 Broken &  -2.7 $\pm$ 0.6\\
 Binned &  -6.1$\pm$ 0.6\\
  Lasenby \& Doran & 4.1 $\pm$ 0.5\\
    \hline
\end{tabular}
\label{table6}
\end{center}
\end{table}

\section{Conclusions}
We have performed a full Bayesian analysis of various parameterisations of the primordial power spectrum which includes not only  the
estimation of cosmological and spectral parameters but also a value of the Bayesian evidence.  This method of model selection was tested
using simulated current and future data, though it is difficult to make conclusive determinations in most cases using real data at
present. 
Consistently high evidence values were obtained for those models incorporating a reduction in power at low $k$, a natural result given the
1$^{st}$ year WMAP data. In particular we found strong evidence in favour of
the Lasenby \& Doran spectrum within a limited MCMC analysis.

\section*{Acknowledgements}
We acknowledge A. Slosar for permission to use his Bayesian evidence code, 
S. Bridle, A. Lewis and A. Liddle for useful discussions. This work was carried
out largely on the COSMOS UK National Cosmology Supercomputer at DAMTP,
Cambridge and we would like to thank S. Rankin and V. Treviso for their
computational assistance. 
MB was supported by a Benefactors Scholarship at St. John's College, Cambridge and an Isaac
Newton Studentship.

\appendix

\label{lastpage}


\begin{thebibliography}{99}


\bibitem[\protect\citeauthoryear{Abazajian et al.}{2003}]{Abazajian}
Abazajian K., et al., 2003, ApJ, 126, 2081
\bibitem[\protect\citeauthoryear{Adams, Ross \& Sarkar}{Adams et al.}{1997}]{Adams}
Adams J.A., Ross G.G. \& Sarkar S., 1997, Nucl. Phys. B. 503, 405
\bibitem[\protect\citeauthoryear{Barriga et al.}{2001}]{barriga}
Barriga J., Gaztanaga E., Santos M.G., Sarkar S., 2001, MNRAS, 324, 977 
\bibitem[\protect\citeauthoryear{Beltran et al.}{2005}]{Beltran}
Beltran M., Garcia-Bellido J., Lesgourgues J., Liddle A., Slosar A., 2005, 
Phys. Rev. D, 71, 063532 
\bibitem[\protect\citeauthoryear{Bridle et al.}{2003}]{Recon}
Bridle S., Lewis A., Weller J., Efstathiou G., 2003, MNRAS, 342, L72
\bibitem[\protect\citeauthoryear{Contaldi et al.}{2003}]{Contaldi}
Contaldi C.R., Peloso M., Kofman L., Linde A., 2003, JCAP 0307
\bibitem[\protect\citeauthoryear{Dickinson et al.}{2004}]{VSA}
Dickinson C. et al., 2004, MNRAS, 353, 732
\bibitem[\protect\citeauthoryear{Dodelson}{2002}]{Dodelson}
Dodelson S., Stewart E., 2002, Phys. Rev. D. 65, 101301
\bibitem[\protect\citeauthoryear{Efstathiou}{2003a}]{efstathioua}
Efstathiou G., 2003, MNRAS, 346, 26
\bibitem[\protect\citeauthoryear{Efstathiou}{2003b}]{efstathioub}
Efstathiou G., 2003, MNRAS, 343 L95 
\bibitem[\protect\citeauthoryear{Freedman et al.}{2001}]{Freedman}
Freedman W.L., et al., 2001, ApJ, 553, 47
\bibitem[\protect\citeauthoryear{Guth}{1981}]{guth:1981}
Guth A., 1981, Phys. Rev. D, 23, 347
\bibitem[\protect\citeauthoryear{Hannestad}{2004}]{Steen}
Hannestad S., 2004, J. Cosmol. Astropart. Phys., JCAP 04(2004)002
\bibitem[\protect\citeauthoryear{Hinshaw et al.}{2003}]{WMAP2}
Hinshaw G. et al., 2003, Astrophys. J. Suppl., 148, 135
\bibitem[\protect\citeauthoryear{Hobson \& McLachlan}{2003}]{evidence}
Hobson M.P., McLachlan C., 2003, MNRAS, 338, 765
\bibitem[\protect\citeauthoryear{Jeffreys}{1961}]{Jeffreys}
Jeffreys H., 1961, \emph{Theory of Probability}, 3rd ed., Oxford University Press
\bibitem[\protect\citeauthoryear{Kogut et al.}{2003}]{WMAP3}
Kogut A. et al., 2003, Astrophys. J. Suppl., 148, 161
\bibitem[\protect\citeauthoryear{Kuo et al.}{2004}]{ACBAR}
Kuo C.L. et al., 2004, Ap. J., 600, 32
\bibitem[\protect\citeauthoryear{Lasenby \& Doran}{2005}]{Doran}
Lasenby A.N., Doran, C., 2005, Phys.Rev. D 71, 063502
\bibitem[\protect\citeauthoryear{Lewis \& Bridle}{2002}]{cosmomc}
Lewis A. and Bridle S., 2002, Phys. Rev. D, 66, 103511
\bibitem[\protect\citeauthoryear{Lewis, Challinor \& Lasenby}{Lewis et al.}{2000}]{camb}
Lewis A., Challinor A., Lasenby A., 2000, ApJ, 538, 473
\bibitem[\protect\citeauthoryear{Linde}{1983}]{Linde}
Linde A.D., 1983, Phys. Rev. Lett. B, 129, 177
\bibitem[\protect\citeauthoryear{Lyth and Riotto}{1998}]{Lyth}
Lyth D.H., Riotto A., 1999, Phys.Rept. 314 1-146
\bibitem[\protect\citeauthoryear{Mukherjee, Parkinson \& Liddle}{Mukherjee et al.}{2005}]{Parkinson}
Mukherjee, P. Parkinson D. Liddle, A., 2005, astro-ph/0508461
\bibitem[\protect\citeauthoryear{Mukherjee and Wang}{2003}]{Wang}
Mukherjee P. \& Wang Y., 2003, Astrophys. J., 593, 38
\bibitem[\protect\citeauthoryear{Niarchou et al.}{2004}]{Niarchou}
Niarchou A., Jaffe A., Pogosian L., 2004, Phys.Rev. D 69 063515
\bibitem[\protect\citeauthoryear{Percival et al.}{2001}]{2dF}
Percival W.J. et al., 2001, MNRAS, 327, 1297
\bibitem[\protect\citeauthoryear{Readhead et al.}{2004}]{CBI}
Readhead A.C.S. et al., 2004, ApJ, 609, 498--512
\bibitem[\protect\citeauthoryear{Rebolo et al.}{2004}]{Rebolo}
Rebolo R. et al., 2004, MNRAS, 353, 747
\bibitem[\protect\citeauthoryear{Slosar \& Hobson}{2003}]{COG}
Slosar A. and Hobson M.P., 2003, astro-ph/0307219
\bibitem[\protect\citeauthoryear{Slosar et al.}{2003}]{Slosar}
Slosar A. et al., 2003, MNRAS, 341, L29
\bibitem[\protect\citeauthoryear{Shafieloo \& Souradeep}{2003}]{Souradeep}
Shafieloo, A., Souradeep, T., 2004, Phys. Rev. D, 70, 043523 
\bibitem[\protect\citeauthoryear{Spergel et al.}{2003}]{Spergel}
Spergel D.N. et al., 2003, Astrophys. J. Suppl., 148, 175
\bibitem[\protect\citeauthoryear{Tocchini-Valentini, Douspis \& Silk}{Tocchini-Valentini et al.}{2005}]{Silk}
Tocchini-Valentini D., Douspis M., Silk J., 2005, MNRAS 359
\bibitem[\protect\citeauthoryear{Trotta}{2005}]{Trotta}
Trotta R., 2005, astro-ph/0504022
\bibitem[\protect\citeauthoryear{Wang}{1994}]{Wang}
Wang Y., 1994, Phys. Rev. D, 50, 6135
\bibitem[\protect\citeauthoryear{Verde et al.}{2003}]{WMAP}
Verde L. et al., 2003, Astrophys. J. Suppl., 148, 195
\end{thebibliography}
\end{document}